\documentclass[conference]{IEEEtran}
\IEEEoverridecommandlockouts
\usepackage{cite}
\usepackage{amsmath,amssymb,amsfonts}
\usepackage{algorithmic}
\usepackage{graphicx}
\usepackage{textcomp}
\usepackage{xcolor}
\usepackage{url}
\usepackage{booktabs}
\usepackage{multirow}
\usepackage{tcolorbox}
\newenvironment{mybox}{\begin{tcolorbox}}{\end{tcolorbox}}
\usepackage{wasysym}
\usepackage{marvosym}

\def\BibTeX{{\rm B\kern-.05em{\sc i\kern-.025em b}\kern-.08em
    T\kern-.1667em\lower.7ex\hbox{E}\kern-.125emX}}

\newcommand{\sysname}{{MARD}}

\title{MARD: A Multi-Agent Framework for Robust Android Malware Detection}

\author{
Xueying Zeng$^{\star}$\textsuperscript{$\S$}\thanks{\textsuperscript{$\S$} Equal contribution.},  
Youquan Xian$^{\ddagger}$\textsuperscript{$\S$},
Sihao Liu$^{\star}$, 
Xudong Mou$^{\star}$,
Yanze Li$^{\star}$, 
Lei Cui$^{\ddagger}$,
Bo Li$^{\star}$\textsuperscript{\Letter}\thanks{\textsuperscript{\Letter} Corresponding author.}
\\
\\
$^{\star}$ School of Computer Science and Engineering, Beihang University, Beijing, China \\
$^{\ddagger}$ School of Cyberspace Security, Beijing University of Posts and Telecommunications, Beijing, China

}

\begin{document}
\maketitle

\begin{abstract}

With the rapid evolution of Android applications, traditional machine learning-based detection models suffer from concept drift. Additionally, they are constrained by shallow features, lacking deep semantic understanding and interpretability of decisions. Although Large Language Models (LLMs) demonstrate remarkable semantic reasoning capabilities, directly processing massive raw code incurs prohibitive token overhead. Moreover, this approach fails to fully unleash the deep logical reasoning potential of LLMs within complex contexts. To address these limitations, we propose MARD, a \underline{m}ulti-\underline{a}gent framework for \underline{r}obust Android malware \underline{d}etection. This framework effectively bridges the gap between the semantic understanding of LLMs and traditional static analysis. It treats underlying deterministic analysis engines as on-demand execution tools, while utilizing the LLM to orchestrate the entire decision-making process. By designing an autonomous multi-agent interaction mechanism based on the ReAct paradigm, MARD constructs a highly interpretable evidentiary chain for conviction. Furthermore, we radically reduce the total cost of conducting a deep analysis of a single complex APK to under \$0.10. Evaluations demonstrate that, without any domain-specific fine-tuning, MARD achieves an F1 score of 93.46\%. It not only outperforms continual learning baselines but also exhibits robustness against concept drift and strong cross-domain generalization capabilities in evaluations spanning up to five years.
\end{abstract}


\section{Introduction}
The Android operating system occupies a dominant position in the mobile ecosystem, making it the primary target for global malware attacks and posing a severe security threat to hundreds of millions of users for an extended period \cite{faruki2014android, bhat2019survey}. Consequently, both academia and industry are continuously committed to enhancing malware detection capabilities \cite{burguera2011crowdroid, yuan2014droid, arp2014drebin, jordaney2017transcend, barbero2022transcending, qiu2022cyber}. Early research primarily relied on extracting static features such as API calls \cite{aafer2013droidapiminer, alazab2020intelligent, chen2022cruparamer}, permissions \cite{li2018significant, thiyagarajan2020improved}, and function call graphs \cite{wu2019malscan, cai2021learning, li2022dmalnet}. These approaches utilized data-driven methods based on Machine Learning (ML) or Deep Learning (DL) to fit statistical patterns of malicious behavior on large-scale training sets. However, with the rapid iteration of the Android ecosystem, the distribution of benign and malicious applications in shallow feature spaces has undergone significant shifts \cite{tang2023demystifying}. This causes detection models trained on historical data to face severe concept drift and performance aging issues upon deployment \cite{guerra2024machine}.

To maintain long-term detection efficacy in dynamically evolving environments, MaMaDroid \cite{onwuzurike2019mamadroid} introduces an API call abstraction mechanism and utilizes Markov chains to model call sequences and capture program behavioral patterns. Subsequently, some research perspectives delved into the semantic level \cite{xu2020sdac, zhang2022slowing, yang2024novel}. These methods effectively capture the invariant core behavioral rules of malware by extracting the semantic features of APIs and mapping newly emerged APIs into existing semantic spaces. Meanwhile, LDCDroid \cite{liu2025ldcdroid}, CADE \cite{yang2021cade}, and FeSAD \cite{fernando2024fesad} approach the problem from the perspective of data distribution. They address the distribution shifts caused by malware evolution by deeply learning and capturing data drift characteristics. To reduce the retraining costs associated with model adaptation to evolution, Xu et al. \cite{xu2019droidevolver} and Huang et al. \cite{huang2024strengthening} explored automatic incremental update mechanisms based on pseudo-labeling techniques. Furthermore, Chen et al. \cite{chen2023continuous} combined active learning with similarity uncertainty sampling in a continual learning approach. This method effectively overcomes the rapid failure of detection models by precisely selecting high-value new samples for manual annotation and model retraining.

Although existing continual learning and active learning strategies alleviate model aging to some extent, traditional deep learning methods still possess inherent limitations. First, their mining of deep semantic information, such as the contextual logic of API calls, permission configurations, and their intentional correlation with real behaviors, remains insufficient. Second, such models typically operate as black boxes, lacking interpretability of decisions, and still rely on continuous data annotation and retraining to maintain performance. In recent years, Large Language Models (LLMs) have made significant progress, and their semantic understanding and reasoning capabilities in complex contexts have improved rapidly \cite{wei2025plangenllms}. SRDC \cite{zhou2025srdc}, AppPoet \cite{zhao2025apppoet}, ForeDroid \cite{li2025foredroid}, and Yan et al. \cite{yan2025prompt} utilized LLMs to assist traditional classification models in semantic enhancement at the feature dimension. Simultaneously, Zheng et al. \cite{zheng2025av} combined ML outputs with LLMs for comprehensive reasoning. Furthermore, He et al. \cite{he2025benchmarking} and Qian et al. \cite{qian2025lamd} directly fed decompiled function snippets and code fragments into LLMs for adjudication. However, directly feeding massive and verbose raw code features into LLMs incurs tremendous token consumption. This approach easily hits the context window limits, leading to high deployment costs and low efficiency. Additionally, simply reducing LLMs to advanced feature extractors fails to truly uncover and fully utilize their deep logical reasoning and intent deduction capabilities within complex malware behavior contexts.

To bridge the profound gap between the deep semantic reasoning of LLMs (constrained by context windows) and the precise control-flow tracking of static analysis (inherently lacking intent awareness), this paper proposes {\sysname}, a multi-agent framework for robust Android malware detection. Specifically, deterministic underlying static analysis engines (such as Soot and FlowDroid) serve as on-demand execution tools, while the LLM acts as the orchestrating brain. Together, they construct an end-to-end collaborative architecture that autonomously executes macro-level screening, micro-level forensics, and global adjudication.

The main contributions of this paper are as follows:
\begin{itemize}
\item We propose {\sysname}, which deeply integrates the continuously escalating high-order semantic reasoning capabilities of LLMs with underlying static analysis engines. It constructs an Android malware detection framework capable of executing high-dimensional intent reasoning and low-level logical forensics without any domain-specific fine-tuning.
\item We design a multi-agent autonomous interaction mechanism based on the ReAct paradigm. This mechanism autonomously plans actions for high-risk APIs and dynamically drives the underlying engines to extract definitive control-flow and data-flow slices, realizing an interpretable evidentiary chain for conviction.
\item By leveraging a heterogeneous model strategy, {\sysname} achieves precise allocation of computational resources, compressing the total end-to-end cost of deeply analyzing a single complex APK to under \$0.10.
\item Extensive empirical evaluations demonstrate that {\sysname} achieves an F1-score of 93.46\% on the datasets. It comprehensively outperforms continual learning baseline models reliant on large-scale data training. Furthermore, in long-span tests from 2017 to 2021, the framework exhibits profound resilience against concept drift and robust cross-domain generalization capabilities.
\end{itemize}

The remainder of this paper is structured as follows:
Section \ref{bg} introduces the relevant background knowledge and existing challenges. Section \ref{main} briefly outlines the framework of {\sysname}.
Section \ref{method} provides the implementation details of the proposed scheme.
Section \ref{experiment} presents the experimental results and analysis. Finally, Section \ref{conclusion} concludes this paper and envisions future work.

\section{Related Work}
\label{bg}
\subsection{Learning-based Malware Analysis}

API calls, as the core static features characterizing the key behaviors of applications, are widely used in Android malware detection systems \cite{cui2023api2vec, cui2024apibeh}. AppContext \cite{yang2015appcontext} utilizes static analysis to extract the contextual features of sensitive APIs in applications, including their triggering events and related control-flow factors. Yumlembam et al. \cite{yumlembam2023android} captured the differences in API usage patterns between benign and malicious applications by quantifying a score for each API. Beyond focusing solely on the critical feature of APIs, the methods in \cite{kim2018multimodal, cai2021jowmdroid, qiu2022cyber, liu2023mobipcr} integrate multiple features to construct sequential representations. These approaches combine these representations with ML or DL techniques, further improving malware classification accuracy.

However, both the Android ecosystem and malware itself continuously evolve. This evolution results in feature discrepancies between old and new applications. Consequently, detection models trained on historical data experience a significant decline in performance when encountering novel malware. This problem is known as concept drift or model aging. To maintain long-term detection capabilities, MaMaDroid \cite{onwuzurike2019mamadroid} constructs static methods resilient to API changes. Subsequently, frameworks like SDAC \cite{xu2020sdac}, APIGRAPH \cite{zhang2022slowing}, and AMDASE \cite{yang2024novel} extract semantic embeddings of APIs. They conduct clustering analysis to capture the invariant essential behavioral patterns of malware at the feature level. This strategy effectively delays model aging. In addition to mining invariant underlying features, another line of research focuses on adapting to changes in data distribution. Yang et al. \cite{yang2021cade} pioneered the use of contrastive learning to detect and explain concept drift samples in security applications within the latent space. Following this trajectory, FeSAD \cite{fernando2024fesad, fernando2022fesa} constructs dedicated drift adaptation layers for ransomware and Android malware. These layers proactively perceive and quantify the distribution shifts caused by evolution. Regarding dynamic model updating, DroidEvolver \cite{xu2019droidevolver} and LDCDroid \cite{liu2025ldcdroid} utilize pseudo-labels to update or retrain detection models by dynamically identifying data drift in new applications. However, the presence of noise in these pseudo-labels can lead to rapid deterioration of model performance. To address this, Chen et al. \cite{chen2023continuous} introduced a sample selection strategy from active learning. Their work proposes a continual learning mechanism to tackle continuously evolving malware more reliably.

Nevertheless, all the aforementioned data-driven paradigms fundamentally remain constrained by the shallow fitting of statistical features. They lack the semantic understanding capability required to grasp the deep business intent and complex code logic of applications. Moreover, frequent model retraining in continual or active learning consumes massive computational resources and incurs high expert annotation costs. Furthermore, the inherent black-box nature of deep learning renders these models incapable of outputting a definitive, code-level evidentiary chain for conviction.

\begin{figure}[htbp]
\centering
\includegraphics[width=0.98\linewidth]{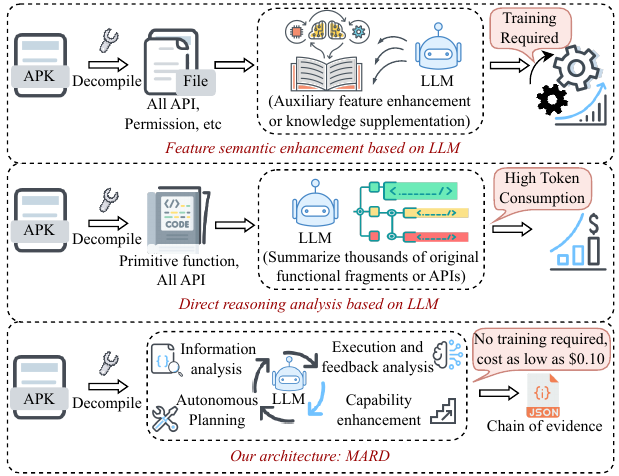}
\caption{Comparison of different scheme architectures.}
\label{fig:Scheme comparison}
\end{figure}

\subsection{LLM-based Malware Analysis}
Large Language Models, with their exceptional code comprehension and commonsense reasoning capabilities, offer a highly promising technological pathway for addressing cybersecurity challenges \cite{wang2024rethinking}. The utilization of LLMs to assist in Android malware analysis can be broadly categorized into two main paradigms: LLM-based feature semantic enhancement and LLM-based direct reasoning analysis. The specific architectural differences are illustrated in Fig. \ref{fig:Scheme comparison}.

In the feature semantic enhancement paradigm, there is a tendency to treat the LLM as an advanced feature extractor to compensate for the deficiencies of traditional machine learning models in semantic understanding. AppPoet \cite{zhao2025apppoet} prompts the LLM to generate natural language summaries for three distinct views: permissions, APIs, and URLs \& Uses-features. These summaries are subsequently fed into a Deep Neural Network (DNN) for downstream classification. Yan et al. \cite{yan2025prompt} and Li et al. \cite{li2025foredroid} leveraged LLMs to translate obscure API call sequences or code invocation chains into descriptive text enriched with domain knowledge. CNN-based or unsupervised learning models then process this text to perform anomaly detection. Furthermore, SRDC \cite{zhou2025srdc} utilizes a ransomware semantic knowledge base, collaboratively constructed by human experts and LLMs, to pre-train a GPT-2 model. This pre-training enhances the model's capability to capture the semantics of zero-day attacks.

Conversely, the direct reasoning analysis paradigm attempts to stimulate and harness the autonomous decision-making and end-to-end adjudication potential of LLMs. AV-Agent \cite{zheng2025av} integrates the confidence scores of ML models and key string features into two distinct reasoning stages of the LLM to yield the final analysis result. LLMalware \cite{ma2025poster} utilizes LLMs to achieve automated feature extraction and fusion. It mitigates the concept drift problem by dynamically updating an external knowledge base. At a more fine-grained level, CAMA \cite{he2025benchmarking} delves into the function level. It requires the LLM to output function summaries and quantified maliciousness scores to evaluate localized malicious intent. LAMD \cite{qian2025lamd} proposes a progressive framework starting from predefined suspicious APIs. It sequentially feeds the extracted structured knowledge into the LLM according to a Function-API-APK hierarchical structure to make the final adjudication.

Despite the significant progress achieved by the aforementioned explorations, some studies remain highly dependent on downstream classifiers. This dependency constrains the system with exorbitant model retraining overheads. Furthermore, directly feeding the full volume of code, invocation chains, or massive API contexts into LLMs easily triggers context window truncation and incurs exceedingly high token overhead. Most crucially, some methods treat LLMs merely as feature summarization tools. They fail to truly unleash the immense potential of LLMs for deep logical reasoning within complex, malicious behaviors.

\section{{\sysname} Overview}
\label{main}

\begin{figure*}[htbp]
\centering
\includegraphics[width=0.95\linewidth]{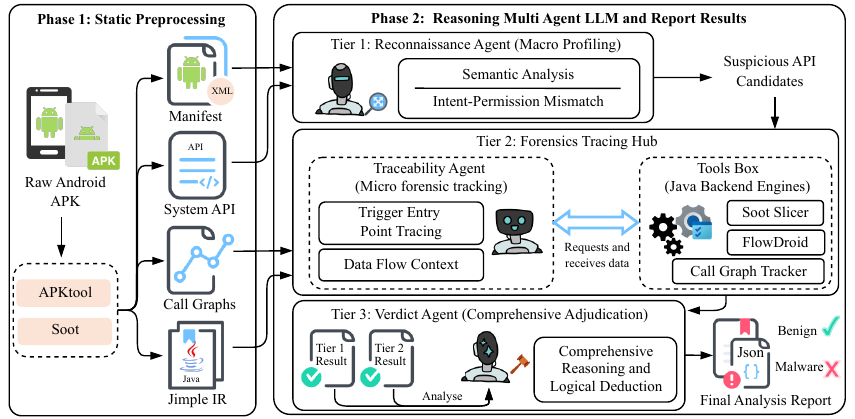}
\caption{Overview of the {\sysname} Architecture.}
\label{fig:MARD}
\end{figure*}

Data-flow-based static analysis tools provide high determinism and precise instruction-level tracking capabilities, yet they fundamentally lack an understanding of high-level malicious semantics and the true intent of developers. Conversely, LLMs exhibit remarkable emergent capabilities in zero-shot semantic reasoning. However, their direct application to massive raw Android Application Packages (APKs) is entirely infeasible due to strict context window constraints. Consequently, we propose {\sysname}, a multi-agent framework for robust Android malware detection. As illustrated in Fig. \ref{fig:MARD}, {\sysname} deeply integrates the continuously escalating high-order semantic reasoning capabilities of LLMs with underlying static analysis engines. This integration constructs an architecture capable of executing high-dimensional intent reasoning and low-level logical forensics without requiring any domain-specific fine-tuning. Our architecture consists of four core modules:

\textbf{Deterministic Static Representation}. Serving as the data foundation of the system, this module leverages the Apktool and Soot engines to perform reverse engineering and dimensionality reduction on the APK. It extracts critical configuration metadata from \texttt{AndroidManifest.xml}. Furthermore, it elevates obscure Dalvik bytecode into the Jimple Intermediate Representation (IR), which closely aligns with human semantics. Additionally, it pre-constructs a global Call Graph (CG) and an API inverted index.

\textbf{Macro-level Heuristic Screening}. The Reconnaissance Agent conducts a review of the Manifest through cross-modal semantic analysis. This process precisely identifies semantic misalignments between the application's declared intent and its requested permissions. Subsequently, it utilizes these macro-level warnings to perform extensive filtering across the global API index. This action exponentially condenses the vast search space into a minimal set of high-risk API candidates. Consequently, it significantly alleviates the cognitive load and token overhead for subsequent deep analysis.

\textbf{Micro-level Autonomous Forensics}. Acting as the definitive forensic hub of the system, the Traceability Agent dynamically drives the underlying advanced static analysis tools. Targeting the high-risk API candidates, the agent autonomously plans analytical actions. It commands the backend engines to execute backward call graph traversal to determine the triggering source. Moreover, it issues taint analysis and program slicing instructions to trace the precise data-flow context. After multiple iterations, fragmented code snippets are abstracted into conclusive evidence vectors with strict topological constraints.

\textbf{Multi-dimensional Evidence Fusion}. The Verdict Agent, situated at the apex of the architecture, is responsible for completing the global logical closed loop. It deeply fuses the macro-level intentional risks extracted in Tier 1 with the structured, micro-level definitive evidence vectors from Tier 2. Based on this fusion, it executes high-order logical deduction and alignment with threat modeling. Ultimately, this agent outputs a standardized JSON analysis report.

\section{Proposed Methodology}
\label{method}
This section details the overall design and implementation of the {\sysname} framework. By constructing a reduced-dimensional data foundation through deterministic analysis engines and relying on multi-agent collaboration to execute high-order semantic reasoning, this framework achieves end-to-end autonomous analysis and adjudication, bridging syntax-level program features to intent-level security semantics.

\subsection{LLM-Oriented Deterministic Code Representation and Dimensionality Reduction}
Android applications typically contain millions of lines of highly obfuscated Dalvik bytecode. To transform this convoluted binary logic into structured knowledge computable by LLMs, we construct a deterministic data foundation.

Given a target application $\mathcal{A}$, we first extract its global configuration manifest $\mathcal{M}$ through reverse engineering. For the core logic code, we utilize the Soot engine to disassemble and elevate the Dalvik bytecode into the Jimple IR. As a strongly-typed three-address code, Jimple abstracts away register allocation and low-level execution details. Simultaneously, it preserves control-flow semantics close to the source code level. This abstraction significantly reduces the complexity of program analysis.

Building upon this, we perform fine-grained syntax tree parsing on the Jimple IR to extract the comprehensive set of API calls $\mathcal{P}$. Furthermore, we introduce the Android system API set $\mathcal{W}$ (built upon the Android SDK API Level 36) to perform semantic filtering on $\mathcal{P}$. This step removes third-party library and framework calls, achieving noise suppression and search space compression. Ultimately, we construct the application's global call graph $G = (V, E)$, where nodes $V$ represent methods and edges $E$ denote invocation dependencies. In addition, we build an API inverted index space with $\mathcal{O}(1)$ time complexity. This index precisely maps each suspicious API signature to specific nodes and code lines in the abstract syntax tree. Consequently, it provides exact coordinates for the subsequent micro-level navigation of the agents.

\subsection{Intent-Permission Semantic Alignment Based on Zero-Shot Reasoning}
\label{Intent permission alignment}
After completing the syntax-level dimensionality reduction, the system's Reconnaissance Agent first intervenes at the macro level. This stage simulates the heuristic screening process of security experts. It rapidly locates potential attack surfaces by reasoning about the semantic consistency between the application's declared intent and its permission requests.

\begin{figure}[htbp]
\centering
\includegraphics[width=0.95\linewidth]{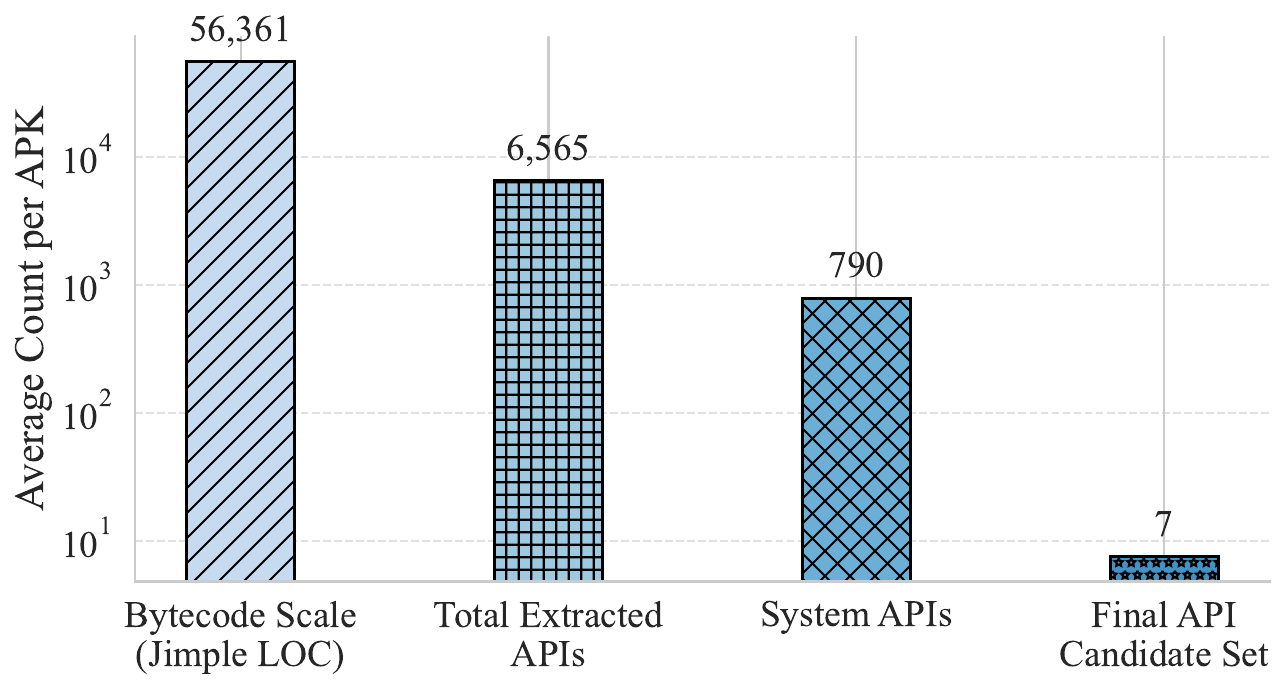}
\caption{Dimension Reduction of API Context per APK.}
\label{fig:API_Context_Funnel}
\end{figure}

Specifically, the Reconnaissance Agent takes the manifest file $\mathcal{M}$ as input. It leverages the zero-shot commonsense reasoning capabilities of the LLM to extract the application's declared intent set $I_{decl}$, and the actually requested sensitive permission set $P_{req}$. Internally, the agent implicitly evaluates the semantic consistency between $I_{decl}$ and $P_{req}$ based on the LLM's prior knowledge. When a discrepancy between the application's functional scope and its requested permissions triggers a structured risk signal, it indicates a potential privilege overreach. For instance, a simple tool application requesting an unreasonable $\texttt{READ\_SMS}$ permission would be flagged as a high-risk anomaly. Based on these macro-level warnings, the Reconnaissance Agent performs mapping within the pre-constructed API inverted index. It prunes the raw API set $\mathcal{P}$ into a minimal, high-risk API candidate subset $\mathcal{C}_{sus}$. As illustrated in Fig. \ref{fig:API_Context_Funnel}, this process significantly condenses the search space and effectively minimizes the token overhead of LLM reasoning.

\subsection{Tool-Augmented Autonomous Context Traceability}\label{Context traceability}

\begin{figure*}[htbp]
\centering
\includegraphics[width=0.95\linewidth]{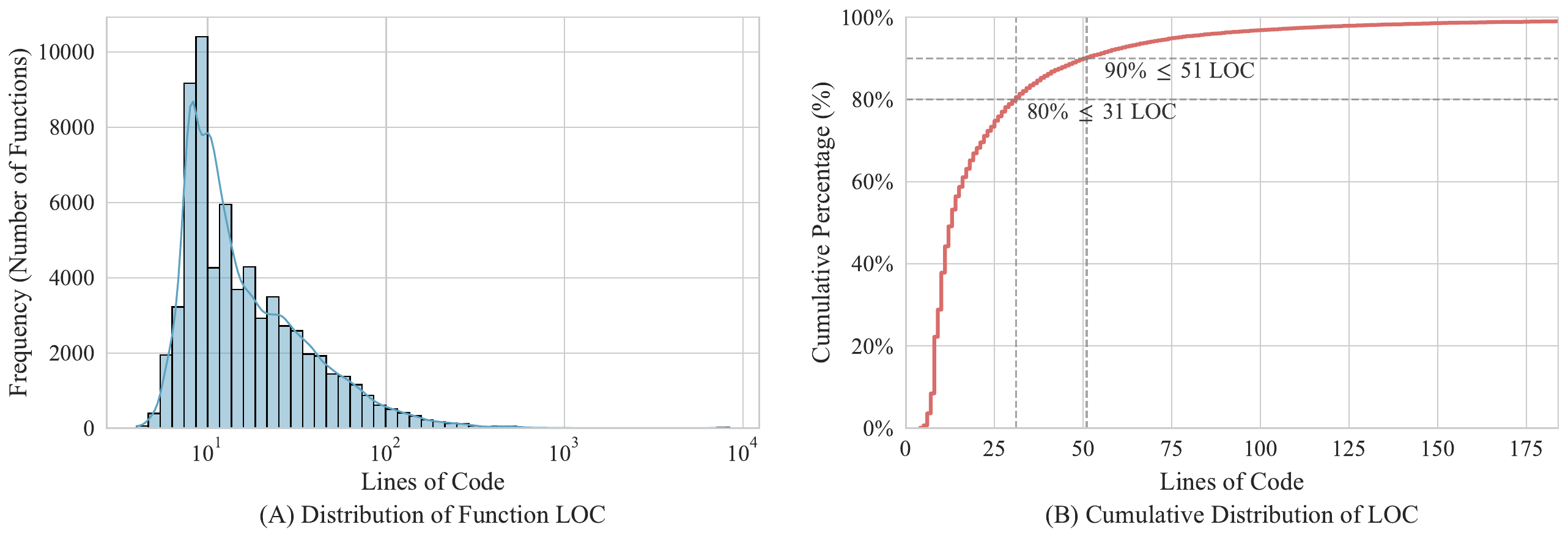}
\caption{Distribution and cumulative distribution of function-level lines of code (LOC). The left subfigure show the histogram and kernel density estimation of LOC, revealing a long-tailed distribution. The right subfigure presents the cumulative distribution, indicating that approximately 80\% and 90\% of functions contain no more than 31 and 51 LOC, respectively.}
\label{fig:Function_LOC_Analysis}
\end{figure*}

For each suspicious API in the set $\mathcal{C}_{sus}$, we design a Traceability Agent based on the ReAct architecture. Because LLMs cannot endogenously execute program flow computations, this agent is endowed with the authority to invoke underlying Java static analysis engines as tools. This design achieves the decoupling of cognition and computation. For each suspicious API node $v_i \in \mathcal{C}_{sus}$, the agent autonomously plans its forensic path during dynamic interactions. Its action space includes:
\begin{itemize}
\item Local Context Retrieval: To support semantic-sensitive analysis of malicious behavior, we provide fine-grained code navigation capabilities, such as API localization and global search. Leveraging these capabilities, the agent can pinpoint suspicious API calls with statement-level precision and extract highly relevant Jimple Intermediate Representation (IR) snippets. Based on empirical statistics of function-level code size (calculated using APKs from the CICMalDroid 2020 \cite{mahdavifar2020dynamic} dataset), function lengths exhibit a significant long-tail distribution. As shown in Fig. \ref{fig:Function_LOC_Analysis}, approximately 80\% and 90\% of the functions do not exceed 31 and 51 lines of code, respectively. Inspired by this distribution characteristic, we adopt a context window centered on the suspicious API. This window expands by 20 lines before and after, totaling 40 lines. This strategy covers the vast majority of semantic dependencies. Simultaneously, it avoids the introduction of redundant context and noise propagation.

\item Trigger Path Search: The Call Graph engine executes backward call chain traversal and reachability analysis on $G=(V, E)$. This process systematically traces the execution flow from the suspicious API back to the application's entry points. Consequently, the agent determines whether the API is triggered by explicit user interaction (e.g., \texttt{onClick}) or silently awakened in the background by system events (e.g., \texttt{BootReceiver}).

\item Data-Flow Reachability Analysis: The agent invokes FlowDroid to execute deterministic forward taint analysis. It marks the return value of the suspicious API $v_i$ as a taint source (Source) and dynamically verifies whether there exists a reachable path to data sinks (Sink), such as network transmission or local storage.

\item Dependency Slicing Extraction: For critical behaviors such as encryption or external connections, the agent executes backward program slicing. It systematically eliminates irrelevant branches and dead code. Thus, it retains only the minimal dependency subgraph that logically affects the target variables, and filters out deceptive code injections.
\end{itemize}

Through multiple rounds of autonomous Observe-Thought-Action iterations, the Traceability Agent structures the dynamically collected control-flow and data-flow fragments. This process generates a multi-dimensional evidence vector $\mathbf{E}$ with strict topological constraints.

\subsection{Two-Stage Evidence Fusion and Interpretable Adjudication}
\label{Interpretable award}
In the final decision stage, the Verdict Agent acts as the central decision-making hub of the system. It receives the static semantic anomalies $\mathcal{M}_{risk}$ from the macro-level profiling and the dynamic structured evidence vector $\mathbf{E}$ from the micro-level forensics.

During the adjudication process, the agent conducts logical deduction based on the multi-dimensional evidence. It maps isolated code behaviors into high-order malicious semantic patterns. By verifying whether the evidence vector satisfies the behavioral characteristics of specific malware families, it outputs the final classification decision $Y \in \{\text{Benign}, \text{Malicious}\}$. More importantly, unlike the black-box predictions of traditional deep learning models, the Adjudication Agent is capable of outputting fine-grained threat categories and confidence scores. Furthermore, it generates an evidentiary chain fully supported by actual data-flow and control-flow logic, providing a high degree of human readability and interpretability.

\section{Experiments and Result Analysis}
\label{experiment}
\subsection{Experimental Setup}
All experiments were implemented in Python 3.11 and conducted on a machine equipped with an Intel Core i7-12700 processor and 32 GB of RAM. All Large Language Models utilized in this research adopted their official vanilla versions without any task-specific fine-tuning or domain adaptation, including the Qwen, DeepSeek, Gemini, GLM, and GPT series models. Specifically, our \ref{Intent permission alignment} and \ref{Context traceability} modules employ the Qwen3-Coder-30B-A3B-Instruct model, while the \ref{Interpretable award} module utilizes the Gemini-3-Pro model.

\subsubsection{Datasets}
To verify the system's generalization capabilities across in-distribution, cross-temporal, and cross-domain scenarios, we selected three representative Android malware datasets for experimental evaluation: AndroZoo \cite{Allix:2016:ACM:2901739.2903508}, CICMalDroid 2020 \cite{mahdavifar2020dynamic}, and CIC-AndMal2017 \cite{lashkari2018toward}. Specifically, we selected applications from AndroZoo spanning the years 2011 to 2021 and constructed an experimental set based on VirusTotal \footnote {\url{https://www.virustotal.com/}} detection results. Applications flagged by more than 10 engines were treated as malicious samples, while applications with 0 detections were randomly sampled as benign samples. Ultimately, a sub-dataset comprising 14,090 applications was constructed, including 6,968 malicious samples and 7,122 benign samples. For the CICMalDroid 2020 dataset, we utilized a total of 7,825 application samples, consisting of 3,923 malicious samples and 3,902 benign samples. Furthermore, CIC-AndMal2017 was utilized entirely as an unseen test set, primarily designed to challenge and benchmark the models' robustness in cross-domain scenarios.

\subsubsection{Baseline Models}
To comprehensively evaluate the detection performance of {\sysname}, we compared it against four representative categories of baseline methods based on the aforementioned real-world Android application datasets.

\textbf{Static Structural \& Behavioral Models} represent classic static detection methods that rely on deep feature engineering and graph analysis. MaMaDroid \cite{onwuzurike2019mamadroid} is a representative static malware detection system that abstracts API calls to the Package and Family levels, utilizing Markov chains to perform high-dimensional modeling of an application's behavioral transition probabilities. Malscan \cite{wu2019malscan} is a market-oriented malware scanning framework. Its core idea is to introduce centrality analysis techniques from social networks to process and extract key topological features from application structure graphs (e.g., function call graphs).

\textbf{Adaptive \& Drift-Resistant Models} represent adaptive learning methods designed to mitigate software evolution and concept drift issues. DroidEvolver \cite{xu2019droidevolver}, operating as a self-evolving Android malware detection system, utilizes pseudo-labels to automatically update and calibrate its detection model online, thereby coping with the rapid iteration of malware. CL-Malware \cite{chen2023continuous} combines active learning with continual learning mechanisms. It aims to enable the detection model to continuously accumulate knowledge over time, thereby effectively adapting to newly emerging malware variants.

\subsubsection{Evaluation Metrics}Based on the True Positives (TP), True Negatives (TN), False Positives (FP), and False Negatives (FN) from the confusion matrix, we calculate Accuracy (ACC), Precision (Pre), Recall (Rec), and the F1-score (F1). The specific calculations are presented in Table \ref{tab:evaluation metrics}.

\begin{table}[]
\caption{Evaluation Metrics.}
\resizebox{\linewidth}{!}{%
\scriptsize
\begin{tabular}{ll}
\hline
Metric & Formula   \\ \hline
Accuracy & $(\text{TP} + \text{TN})/(\text{TP} + \text{TN} + \text{FP} + \text{FN})$ \\
Precision  & $ (\text{TP})/(\text{TP} + \text{FP})$ \\  
Recall  & $ (\text{TP})/(\text{TP} + \text{FN})$ \\  
F1-score  & $ 2 \cdot (\text{Precision} \cdot \text{Recall})/(\text{Precision} + \text{Recall})$ \\   \hline
\end{tabular}
}
\label{tab:evaluation metrics}
\end{table}

\begin{table*}[htbp]
\centering
\caption{Overall Performance Comparison on Different Datasets (\%)}
\label{tab:overall_performance}
\resizebox{0.95\textwidth}{!}{
\tiny
\begin{tabular}{c c c c c c c c c}
\toprule
\multirow{2}{*}{Method} & \multicolumn{4}{c}{CICMalDroid 2020} & \multicolumn{4}{c}{AndroZoo} \\
\cmidrule(lr){2-5} \cmidrule(lr){6-9}
& Acc & Pre & Rec & F1 & Acc & Pre & Rec & F1 \\
\midrule
MaMaDroid (family)     & 60.16 & 51.04 & 96.08 & 66.67 & 72.58 & 75.40  & 64.30 & 69.34 \\
MaMaDroid (package)    & 48.78 & 44.44 & 94.12 & 60.38 & 48.28 & 48.08 & \textbf{98.67} & 64.62  \\
DroidEvolver      & 40.32 & 40.65 & \underline{98.04} & 57.47 & 69.58 & 74.18 & 50.47 & 57.85 \\
CL-Malware        & \underline{92.74} & \underline{85.00} & \textbf{100.0} & \underline{91.89} & \underline{87.57} & \underline{90.67} & 83.30 & \underline{86.67} \\
Malsca (knn-1)          & 58.00 & 52.94 & 60.00 & 56.25 & 66.26 & 70.48 & 52.90 & 59.76 \\
Malscan (knn-3)         & 44.00 & 42.03 & 64.44 & 50.88 & 62.98 & 64.61 & 47.41 & 53.89 \\
Malscan (random)       & 46.00 & 44.71 & 84.44 & 58.46 & 68.74 & \textbf{91.01} & 40.44 & 54.09 \\
{\sysname} (Ours)                   & \textbf{94.35} & \textbf{89.29} & \underline{98.04} & \textbf{93.46} & \textbf{89.19} & 87.76 & \underline{86.46} & \textbf{87.01} \\
\bottomrule
\end{tabular}
}
\vspace{1ex}

\raggedright
\footnotesize{\textit{\hspace{2em} Note: \textbf{Bold} indicates the best performance, and \underline{underlined} indicates the second-best.}}
\end{table*}

\subsection{Research Questions (RQs)}
\begin{itemize}
\item \textbf{RQ1 (Effectiveness of {\sysname})}: 
How does the proposed zero-shot LLM multi-agent framework perform in Android malware detection compared to traditional learning-based baselines trained on large-scale datasets?

\item \textbf{RQ2 (Temporal Generalization)}:
How resilient is the LLM-based framework against concept drift over time compared to baseline models?

\item \textbf{RQ3 (Cross-Domain Generalization)}: 
Can the proposed framework maintain high detection efficacy across entirely different and unseen dataset distributions without domain-specific retraining?

\item \textbf{RQ4 (Model Capability and Component Analysis)}: 
How do different state-of-the-art LLMs (e.g., GPT-5.2, DeepSeek-R1, GLM-4.7) impact the final verdict accuracy, and what is the impact of removing the micro-level independent forensics module?

\item \textbf{RQ5 (Cost and Feasibility Analysis)}: 
What is the token consumption and economic cost of the multi-agent framework, and how feasible is it for actual deployment?
\end{itemize}

\subsection{Effectiveness of {\sysname} (RQ1)}
We conducted a comprehensive overall performance evaluation of {\sysname} against learning-based Android malware detection baseline models on two heterogeneous datasets: CICMalDroid 2020 and AndroZoo. As shown in Table \ref{tab:overall_performance}, {\sysname} demonstrates outstanding detection efficacy on both datasets. On the CICMalDroid 2020 dataset, {\sysname} achieves the highest accuracy of 94.35\% and an F1-score of 93.46\%. Traditional baselines like DroidEvolver sustain high recall (${\sim}98\%$) at the cost of abysmal precision (${\sim}40\%$) and sub-50\% accuracy, which makes them unusable in practice. Although the State-Of-The-Art (SOTA) continual learning model, CL-Malware, achieves a 100\% recall rate, its precision is only 85.00\%, indicating a tendency to generate a large volume of false positives. In contrast, {\sysname} achieves an optimal balance between a precision of 89.29\% and a recall of 98.04\%.

When the testing environment shifts to the AndroZoo dataset, which is characterized by a more complex data distribution and a significantly larger scale, the performance of traditional data-driven models suffers a catastrophic degradation. The recall rates of MaMaDroid (family) and DroidEvolver plummet from 96.08\% and 98.04\% down to 64.30\% and 50.47\%, respectively. Although Malscan (random) attains a precision of 91.01\% on AndroZoo, its recall is a mere 40.44\%. This implies that it misses nearly 60\% of actual malware, rendering it of very little practical value in real-world security deployments. On the contrary, {\sysname}  does not require any domain-specific fine-tuning, parameter updates, or retraining on the AndroZoo dataset, yet it still maintains an accuracy of 89.19\% and an F1-score of 87.01\%, exceeding the best-performing baseline model, CL-Malware.

\begin{mybox}
    \textbf{Answer for RQ1:} Without any domain-specific training, {\sysname} significantly outperforms SOTA baseline models reliant on large-scale data, achieving an optimal balance between precision and recall on heterogeneous datasets. Our architecture successfully overcomes the catastrophic degradation experienced by traditional data-driven models when confronted with complex samples in dynamic environments.
\end{mybox}

\subsection{Temporal Generalization (RQ2)}
\begin{figure}[htbp]
\centering
\includegraphics[width=0.98\linewidth]{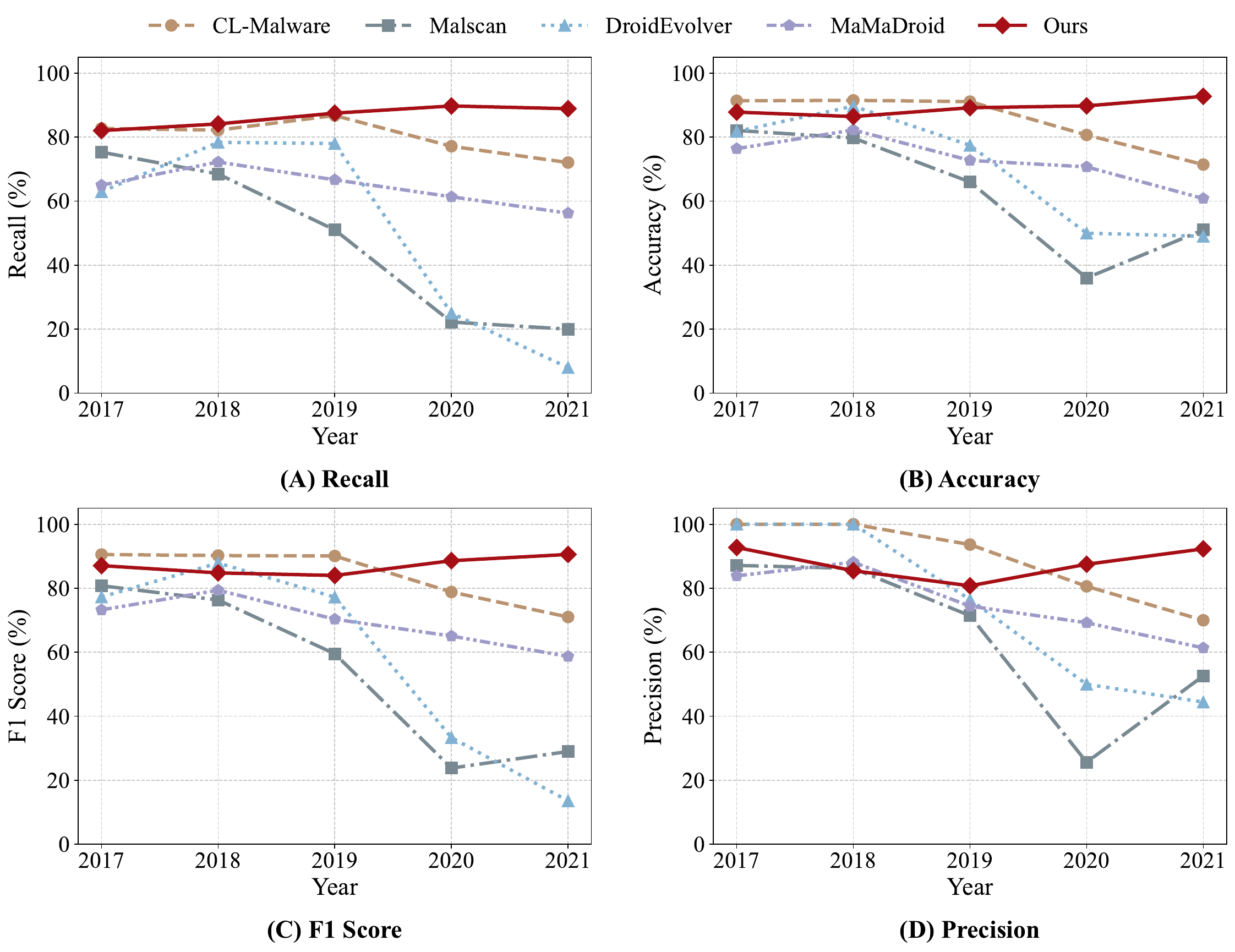}
\caption{2017-2021 results on AndroZoo dataset.}
\label{fig:AndroZoo_year}
\end{figure}

To evaluate the models' temporal generalization capabilities against malware evolution, we adopted a strict chronological evaluation strategy. We partitioned the test set based on the year of the applications' first appearance (from 2017 to 2021). Fig. \ref{fig:AndroZoo_year} illustrates the long-term evolutionary trends of the baseline models and {\sysname}.

The experimental data clearly reveal the catastrophic impact of concept drift on data-driven models. As shown in Fig. \ref{fig:AndroZoo_year}, models relying on fixed feature spaces maintain acceptable performance during 2017-2018, but experience a precipitous drop starting in 2019. Specifically, the recall of DroidEvolver plummets from 78.38\% in 2018 to a mere 8.00\% in 2021, with its F1-score correspondingly plunging to 13.56\%. This implies that when confronted with novel malware variants in 2021, the model almost entirely loses its detection capabilities, degrading to mere random guessing. This exposes the inherent flaw of traditional shallow features, which are highly susceptible to aging as attack vectors evolve.

CL-Malware introduces a dynamic model update mechanism based on active learning to mitigate the aging issue. The data indicate that this strategy delays the performance degradation to some extent, enabling it to maintain an F1-score of approximately 90\% between 2017 and 2019. However, with the evolution of the Android ecosystem post-2020, CL-Malware still fails to overcome the concept drift problem, as its accuracy and F1-score fall to 71.43\% and 71.01\%, respectively, in 2021. This proves that solely performing incremental learning and pseudo-label updating within a legacy feature space cannot fundamentally bridge the gap caused by the high-order semantic mutations of malware.

In stark contrast to all baseline models, {\sysname} demonstrates unparalleled temporal stability and generalization capabilities. Over the five years, the F1-score of {\sysname} remains consistently stable above 84\%. Similarly, its accuracy is maintained at a highly competitive level, stabilizing around 92.75\% in recent years. Also, the recall of {\sysname} remains remarkably steady, within a margin of 82.05\% to 88.89\% throughout the entire timeline. This indicates that the system is profoundly immune to temporal aging, possessing an intrinsic and stable capacity to capture novel threats with complex disguises accurately.

\begin{mybox}
\textbf{Answer for RQ2:} In stark contrast to traditional and continual learning models that suffer from a precipitous performance decline over time, {\sysname} demonstrates exceptional temporal generalization capabilities. By substituting the fitting of shallow statistical features with deep semantic reasoning of attack intent, the framework maintains stable detection efficacy across a time span of up to five years.
\end{mybox}

\subsection{Cross-Domain Generalization (RQ3)}
We constructed a cross-domain evaluation scenario to test the model's cross-domain generalization capabilities on completely unknown datasets. In this setting, all learning-based baseline models underwent large-scale pre-training on the AndroZoo (2011-2016) dataset and were subsequently subjected to direct transfer testing on two target datasets with completely different distributions, namely CICMalDroid 2020 and CIC-AndMal2017.

As shown in Table \ref{tab:CICMalDroid2020_results}, on the CICMalDroid 2020 dataset, the continual learning model CL-Malware achieves the best performance with an F1-score of 95.33\%. This phenomenon indicates that this target dataset might share a high degree of in-distribution overlap in certain underlying statistical features with the baseline models' AndroZoo training set. However, even when facing this extremely advantageous scenario for pre-trained models, {\sysname} still demonstrates astonishing competitiveness, closely following with a high F1-score of 93.46\%. This strongly proves that even in the absence of domain prior knowledge, zero-shot analysis based on multi-agent collaboration can achieve or even rival SOTA models trained on domain-specific data.

\begin{table}[htbp]
\centering
\caption{Cross-Dataset Performance on CICMalDroid 2020 (\%).}
\label{tab:CICMalDroid2020_results}
\resizebox{\linewidth}{!}{%
\begin{tabular}{lcccc}
\toprule
Method & Acc & Pre & Rec & F1 \\
\midrule
MaMaDroid (package) & 57.72 & 40.00 & 3.92 & 7.14 \\
DroidEvolver   & 86.29 & 75.76 & \underline{98.04} & 85.47 \\
CL-Malware     & \textbf{95.97} & \textbf{91.07} & \textbf{100.0} & \textbf{95.33} \\
Malscan (knn-1)            & 83.87 & 73.85 & 94.12 & 82.76 \\
{\sysname} (Ours)                  & \underline{94.35} & \underline{89.29} & \underline{98.04} & \underline{93.46} \\
\bottomrule
\end{tabular}%
}
\end{table}

When the test target shifts to CIC-AndMal2017, as seen in Table \ref{tab:CICAndMal2017_results}, the performance of all data-driven models suffers a catastrophic degradation because the malware family composition, collection strategies, and API usage patterns of this dataset exhibit severe spatial distribution shifts compared to AndroZoo. The F1-score of the most robust model, CL-Malware, shrinks drastically by nearly 21\% (dropping to 74.45\%), and its recall plummets to 64.56\%, indicating that it generates a massive number of false negatives when confronted with unknown domain features. Conversely, {\sysname} demonstrates stability across datasets, achieving an accuracy of 91.14\%.

\begin{table}[htbp]
\centering
\caption{Cross-Dataset Performance on CIC-AndMal2017 (\%).}
\label{tab:CICAndMal2017_results}
\resizebox{\linewidth}{!}{%
\begin{tabular}{lcccc}
\toprule
Method & Acc & Pre & Rec & F1 \\
\midrule
MaMaDroid (package) & 48.30 & 53.33 & 10.39 & 17.39 \\
DroidEvolver   & 68.79 & 67.05 & 74.68 & 70.66 \\
CL-Malware     & \underline{77.71} & \underline{87.93} & 64.56 & 74.45 \\
Malscan (knn-1)     & 74.54 & 70.99 & \underline{83.54} & \underline{76.74} \\
{\sysname} (Ours)                  & \textbf{91.14} & \textbf{92.31} & \textbf{90.00} & \textbf{91.14} \\
\bottomrule
\end{tabular}%
}
\end{table}

\begin{table*}[htbp]
\centering
\caption{Ablation Study on Different LLM Backbones (\%)}
\label{tab:llm_ablation}
\resizebox{0.95\textwidth}{!}{%
\begin{tabular}{ccccccccccccccc}
\toprule
\multirow{2}{*}{Organization} & \multirow{2}{*}{LLM} 
& \multicolumn{4}{c}{CIC-AndMal2017} 
& \multicolumn{4}{c}{CICMalDroid 2020} \\
\cmidrule(lr){3-6} \cmidrule(lr){7-10}
& & Acc & Pre & Rec & F1 
& Acc & Pre & Rec & F1 \\
\midrule
\multirow{2}{*}{Google}
& Gemini-3-pro        & 91.14 & 92.31 & 90.00 & 91.14
 & 94.35 & 89.29 & 98.04 & 93.46 \\
& Gemini-2.5-flash    & 82.91 & 81.18 & 86.25 & 83.64
 & 85.48 & 73.91 & 100.0 & 85.00 \\
\midrule
\multirow{2}{*}{Zhipu}
& GLM-4.7   & 86.71 & 85.54 & 88.75 & 87.12
 & 88.71 & 81.36 & 94.36 & 87.27 \\
 & GLM-4-flash     & 76.43 & 69.37 & 96.25 & 80.63
 & 75.00 & 62.20 & 100.0 & 76.69 \\
\midrule
\multirow{3}{*}{OpenAI}
& GPT-5.2        & 89.87 & 85.56 & 96.25 & 90.59
 & 91.13 & 82.26 & 100.0 & 90.27 \\
& GPT-5-mini         & 82.91 & 80.46 & 87.5 & 83.83
 & 83.87 & 72.46 & 98.04 & 83.33 \\
& GPT-4o         & 81.01 & 75.00 & 93.75 & 83.33
 & 79.84 & 67.11 & 100.0 & 80.31 \\
\midrule
\multicolumn{2}{c}{\textit{w/o Micro-level Autonomous Forensics}} 
& 86.08 & 81.52 & 93.75 & 87.75 
& 83.87 & 71.83 & 100.0 & 83.61 \\
\bottomrule
\end{tabular}
}
\vspace{1ex}

\raggedright
\footnotesize{\textit{\hspace{1em} Note: "w/o Micro-level Autonomous Forensics" represents the system using GPT-5.2 but skipping the code context tracing stage.}}
\end{table*}

\begin{mybox}
\textbf{Answer for RQ3:} In cross-domain evaluation scenarios, when confronted with severe data distribution shifts, traditional learning-based models suffer a precipitous performance decline due to dataset bias. In contrast, {\sysname}, leveraging its evidence-chain-based deep semantic understanding, maintains a detection accuracy exceeding 91\% on completely unseen heterogeneous datasets, demonstrating outstanding and stable cross-domain generalization capabilities.
\end{mybox}

\subsection{Model Capability and Component Analysis (RQ4)}
We investigated the impact across two core dimensions: first, the differences in final adjudication results among Large Language Models of varying capability tiers; second, the necessity of the micro-level autonomous forensics module's existence within the system architecture. The detailed evaluation results are presented in Table \ref{tab:llm_ablation}.

The experimental results clearly indicate that the final detection efficacy of {\sysname} is highly positively correlated with the high-order logical reasoning capabilities of the underlying LLM. Among all evaluated models, the Gemini-3-pro model demonstrates an overwhelming advantage, achieving the best F1-scores of 91.14\% and 93.46\% on CIC-AndMal2017 and CICMalDroid 2020, respectively. Models of the same tier, such as GPT-5.2 and GLM-4.7, also exhibit extremely strong competitiveness. In contrast, models with smaller parameter sizes or those positioned as lightweight, such as Gemini-2.5-flash, GPT-5-mini, and GLM-4-flash, fall into obvious false-positive traps. Taking GLM-4-flash and GPT-4o as examples, their recall on CICMalDroid 2020 both reach 100\%, but their precision plummets to 62.20\% and 67.11\%, respectively. This profoundly reveals that lightweight models suffer from severe cognitive overload and a conservative classification tendency when confronting complex code evidentiary chains; that is, when unable to precisely disentangle data flows, the models tend to uniformly classify all suspicious applications as malicious, thereby generating false positives. Therefore, LLMs equipped with deep reasoning and long-context orchestration capabilities are a necessary prerequisite for achieving precise zero-shot detection.

To verify the necessity of the tool-augmented architecture, we designed an experiment that strips away the micro-level traceability stage (w/o Micro-level Autonomous Forensics). Under this setting, GPT-5.2 can only rely on the Manifest intent misalignment information provided by macro-level heuristic screening for adjudication, lacking the support of actual control-flow and data-flow slices from the underlying Soot/FlowDroid engines. The data in Table \ref{tab:llm_ablation} show that after removing the micro-level autonomous forensics module, the system's precision on CICMalDroid 2020 plummets from 82.26\% to 71.83\%, and the overall F1-score drops to 83.61\%. This precipitous decline intuitively demonstrates the core value of the micro-level autonomous forensics module: macro-level semantic screening is restricted to uncovering potential risks, whereas micro-level autonomous dynamic tool invocation is essential for verifying definitive maliciousness. Lacking the constraint of underlying deterministic code evidence, the LLM falls into unfounded inference, which once again validates the scientific soundness and necessity of the framework proposed in this paper.

\begin{figure*}[htbp]
\centering
\includegraphics[width=0.95\linewidth]{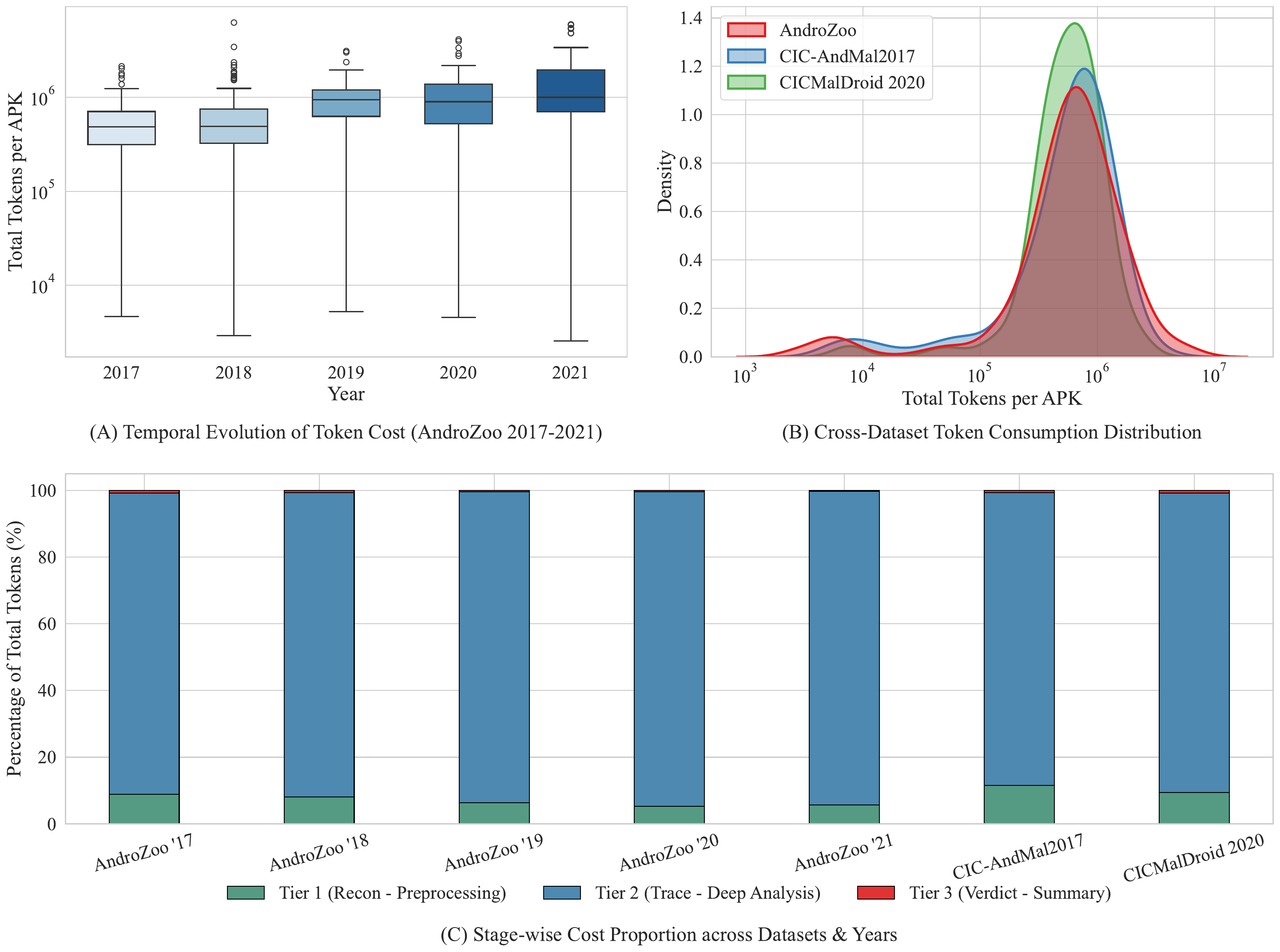}
\caption{Token consumption analysis results.}
\label{fig:Token_Cost}
\end{figure*}

\begin{mybox}
\textbf{Answer for RQ4:} The high-order logical reasoning capabilities of LLMs play a decisive role in system efficacy, and flagship models can effectively avoid the high recall and low precision false-positive trap faced by lightweight models. Furthermore, ablation experiments confirm the indispensability of the micro-level traceability module; removing the underlying toolchain support causes precision to plummet by over 10\%, proving that a deterministic code-level evidentiary chain is the cornerstone for achieving precise adjudication.
\end{mybox}

\subsection{Cost and Feasibility Analysis (RQ5)}
\begin{figure*}[htbp]
\centering
\includegraphics[width=0.95\linewidth]{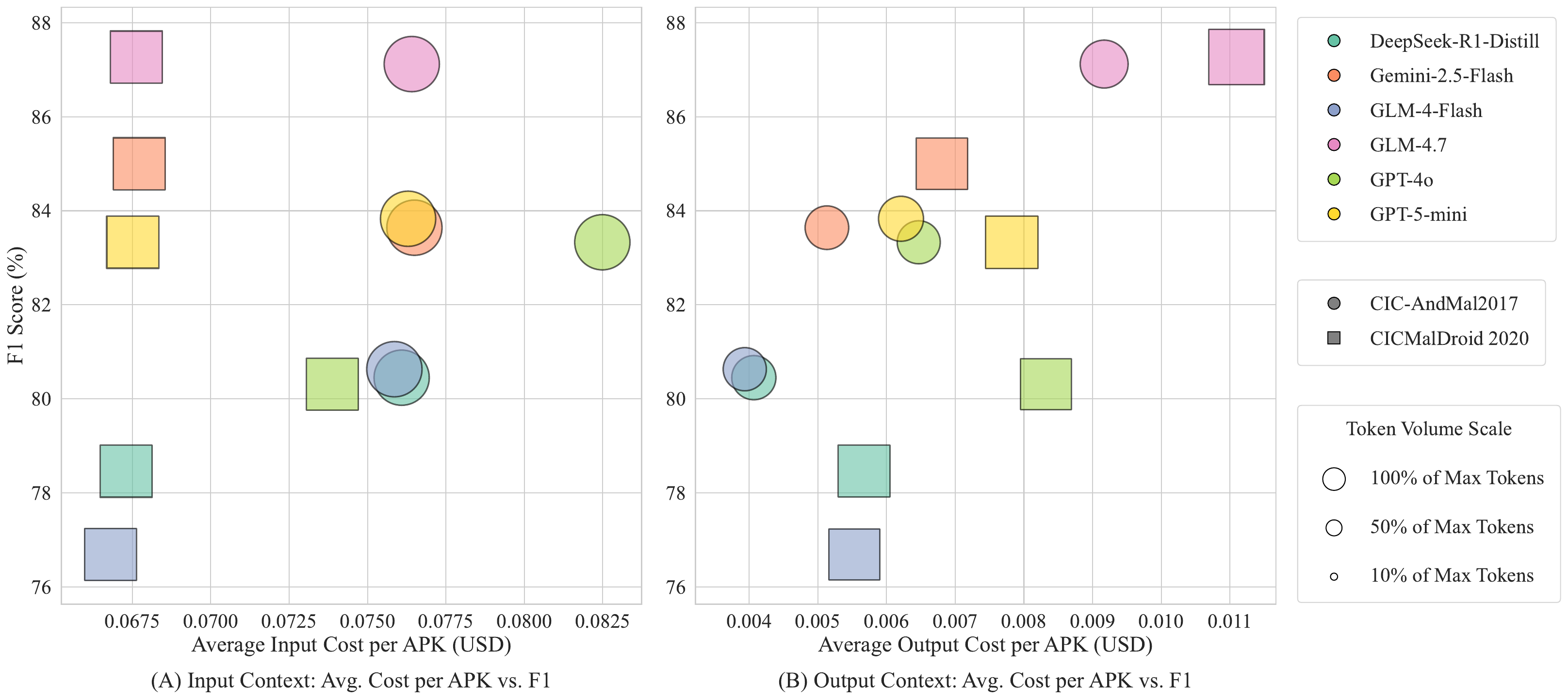}
\caption{Average consumption cost per APK.}
\label{fig:Cost_vs_F1}
\end{figure*}

To comprehensively evaluate the engineering feasibility of {\sysname} in real-world deployment scenarios, we detailed and quantified the token consumption distribution, the cost overhead trends over time, and the economic efficiency of different LLMs when processing a single APK. The evaluation results are illustrated in Fig. \ref{fig:Token_Cost} and Fig. \ref{fig:Cost_vs_F1}.

Fig. \ref{fig:Token_Cost}(C) clearly reveals the internal operational mechanism of the multi-agent framework. Across all test datasets and time spans, Tier 2 (Trace - Deep Analysis) consistently acts as the absolute resource consumption bottleneck, accounting for over 85\% to 90\% of the system's total token consumption. This phenomenon not only aligns with expectations but also strongly validates the core design philosophy of {\sysname}. The macro-level pre-screening in Tier 1 successfully intercepts a massive volume of risk-free code, consuming only a minimal fraction of tokens ($<10\%$); meanwhile, the system skews the vast majority of computational resources towards the dynamic tool interaction (ReAct loop) and deep program slice reading in Tier 2. The token consumption in the Tier 3 adjudication and summary stage is negligible ($<1\%$). This indicates that the LLM is not merely performing superficial text summarization but is substantively engaging in deep code review.

As shown in Fig. \ref{fig:Token_Cost}(A), alongside the increasing complexity of the Android application ecosystem (from 2017 to 2021), the total amount of tokens required to process a single APK exhibits a slow but inevitable upward trend. The Kernel Density Estimation (KDE) curve in Fig. \ref{fig:Token_Cost}(B) further demonstrates that the analysis overhead for the vast majority of APKs is concentrated between $10^5$ and $10^6$ tokens. This highlights the absurdity of strategies that directly feed the full volume of code into an LLM. Without relying on our preliminary static analysis infrastructure for dimensionality reduction, modern giant APKs would easily breach the context limits of any commercial LLM and incur incalculable economic costs. But it is worth noting that our density distribution maintains a consistent shape, ensuring that the cost is stable.

Fig. \ref{fig:Cost_vs_F1} intuitively illustrates the relationship between the average input and output cost (in USD) for processing a single APK and the final detection F1-score. To maximize the feasibility of commercial deployment, {\sysname} adopts an economical heterogeneous model strategy. Specifically, the heavy preliminary tasks, which account for over 90\% of the token consumption in the system, are assigned to the Qwen3-Coder-30B-A3B-Instruct model, which specializes in code comprehension and possesses a significant cost advantage. This establishes a highly affordable foundation for the system. Building upon this, Fig. \ref{fig:Cost_vs_F1} demonstrates the variations in the overall pipeline overhead when switching to different LLMs exclusively during the comprehensive adjudication stage. Because the adjudication stage only needs to process the highly condensed, structured evidence vectors extracted from the first two stages, the context billing for commercial LLMs is drastically reduced. The data indicates that even when utilizing models with exceptionally strong reasoning capabilities, the total input cost for processing a single APK across the entire pipeline is strictly controlled between \$0.0675 and \$0.0825 (Fig. \ref{fig:Cost_vs_F1}(A)), while the output cost is as low as \$0.004 to \$0.011 (Fig. \ref{fig:Cost_vs_F1}(B)). This means that the total overhead for completing a deep, automated code forensics and malware adjudication process is well under \$0.10 \footnote{The Qwen, DeepSeek, and GLM series models were invoked via the API interface provided by SiliconFlow (\url{https://siliconflow.cn/}), while the Gemini and GPT series models were accessed via the API interface provided by OpenRouter (\url{https://openrouter.ai/}).}.

\begin{mybox}
\textbf{Answer for RQ5:} {\sysname} achieves high engineering feasibility and cost efficiency. By using a heterogeneous model architecture, token-intensive preliminary screening is handled by low-cost models, reducing per-APK cost to under \$0.10, while mid-to-large models perform final adjudication to ensure precise and reliable detection, enabling scalable deployment of multi-agent frameworks in industrial settings.
\end{mybox}

\section{Conclusion}
\label{conclusion}
In this paper, we propose {\sysname}, an Android malware detection framework that integrates deep semantic reasoning from LLMs with deterministic static analysis. By leveraging engines like Soot and FlowDroid within a collaborative workflow of macro-level screening, micro-level forensics, and global adjudication, {\sysname} enables zero-shot end-to-end detection with evidentiary chains supported by actual code and data-flow logic. Evaluations over five years of heterogeneous datasets demonstrate strong detection performance with an F1-score of 93.46\%, robust cross-domain generalization, and resilience against concept drift. Furthermore, a heterogeneous model strategy reduces the per-APK analysis cost to under \$0.10. Limitations remain when analyzing heavily packed or dynamically loaded apps. Future work will integrate dynamic sandbox logs and memory forensics to extend {\sysname} toward hybrid static-dynamic analysis for more stealthy threats.


\bibliographystyle{unsrt}
\bibliography{myref}

\end{document}